\pdfoutput=1
\documentclass[aps,prl,twocolumn,superscriptaddress]{revtex4}
\usepackage{txfonts}
\usepackage{graphicx}
\usepackage{color}
\usepackage{fancyhdr} 
\begin{document}

\title{Parity independence of the zero-bias conductance peak in a nanowire based topological superconductor-quantum dot hybrid device}

\author{M. T. Deng}
\affiliation{Division of Solid State Physics, Lund University, Box 118, S-221 00 Lund, Sweden}
\author{C. L. Yu}
\affiliation{Division of Solid State Physics, Lund University, Box 118, S-221 00 Lund, Sweden}
\author{G. Y. Huang}
\affiliation{Division of Solid State Physics, Lund University, Box 118, S-221 00 Lund, Sweden}
\author{M. Larsson}
\affiliation{Division of Solid State Physics, Lund University, Box 118, S-221 00 Lund, Sweden}
\author{P. Caroff}
\affiliation{Division of Solid State Physics, Lund University, Box 118, S-221 00 Lund, Sweden}
\author{H. Q. Xu}
\email[Corresponding author: ]{hongqi.xu@ftf.lth.se}
\affiliation{Division of Solid State Physics, Lund University, Box 118, S-221 00 Lund, Sweden}
\affiliation{Key Laboratory for the Physics and Chemistry of Nanodevices and Department of Electronics, Peking University, Beijing 100871, China}

\date{\today}

\begin{abstract}
We explore the signatures of Majorana fermions in a nanowire based topological superconductor-quantum dot-topological superconductor hybrid device by charge transport measurements. The device is made from an epitaxially grown InSb nanowire with two superconductor Nb contacts on a Si/SiO$_2$ substrate. At low temperatures, a quantum dot is formed in the segment of the InSb nanowire between the two Nb contacts and the two Nb contacted segments of the InSb nanowire show superconductivity due to the proximity effect. At zero magnetic field,  well defined Coulomb diamonds and the Kondo effect are observed in the charge stability diagram measurements in the Coulomb blockade regime of the quantum dot. Under the application of a finite, sufficiently strong magnetic field, a zero-bias conductance peak structure is observed in the same Coulomb blockade regime. It is found that the zero-bias conductance peak is present in many consecutive Coulomb diamonds, irrespective of the even-odd parity of the quasi-particle occupation number in the quantum dot. In addition,  we find that the zero-bias conductance peak is in most cases accompanied by two differential conductance peaks, forming a triple-peak structure, and the separation between the two side peaks in bias voltage shows oscillations closely correlated to the background Coulomb conductance oscillations of the device. The observed zero-bias conductance peak and the associated triple-peak structure are in line with the signatures of Majorana fermion physics in a nanowire based topological superconductor-quantum dot-topological superconductor system, in which the two Majorana bound states adjacent to the quantum dot are hybridized into a pair of quasi-particle states with finite energies and the other two Majorana bound states remain as the zero-energy modes located at the two ends of the entire InSb nanowire.
\end{abstract}


\maketitle

The search for Majorana Fermions~\cite{Majorana} in solid state systems, especially in s-wave superconductor (SC)-coupled semiconductor nanowires (NWs) with a strong spin-orbit interaction (SOI), is one of paramount research tasks in physics today~\cite{Wilczek2009, Franz2010, Service2011, Alicea2012,Nayak2008, ReadGreen, FuKaneSwaveTi,Lutchyn2010, Oreg2010, Stanescu2011}. By exposing an s-wave SC-coupled semiconductor NW with strong SOI to a sufficiently strong and appropriately oriented magnetic field and, thus, driving the system into topological superconductor (TS) phase, zero-energy quasi-particle states, i.e., Majorana Fermions (MFs), are expected to appear in pair at the two ends of the semiconductor NW. Recently, several groups have reported on their observations of the signatures of zero-energy MFs in charge transport measurements of hybrid SC-semiconductor NW devices~\cite{Mourik2012, Deng2012, Das2012, Churchill2013}. In these experiments, InSb or InAs semiconductor NWs are used and contacted by an s-wave SC of NbTiN, Nb, or Al. At zero magnetic field, these superconductor contacted NWs show superconductivity at low temperatures due to the proximity effect. Under the application of an external magnetic field, these NWs are turned to be TS NWs and can host zero-energy MF modes at the ends of the NWs. In the charge transport measurements, these MF states manifest themselves as a zero-bias conductance peak (ZBCP). The quasi-particles carrying these modes are shown theoretically to obey non-Abelian statistics and could be utilized for topological quantum computing. However, the non-Abelian statistics of Majorana quasi-particles in solid state has not been demonstrated experimentally. It has been suggested that an intriguing experimental demonstration of the non-Abelian statistics is to carry out a braiding experiment of two Majorana quasi-particles in a TS system. In such an experiment, the two MF modes from different MF pairs could be brought to interact via a non-topological object. It is, therefore, fundamentally important to study the novel physics of TS systems in the presence of such interaction. 

Several other physical mechanisms could also lead to a ZBCP in the charge transport measurements of the experimentally studied SC-semiconductor NW hybrid systems, such as the Kondo effect, the Josephson supercurrent, Andreev reflections, level crossing, quantum phase transition, transport through impurity states, and the effects of disorder in potential and magnetic field distributions. The most of these effects can be conclusively excluded from the reported experiments. However, the question regarding whether the Kondo correlation could be a physical mechanism for the observed ZBCPs in these experiments or some of these experiments is still in active debate \cite{Lee2012} and requires a conclusive experimental study.

Here, we report on the realization and measurements of a hybrid Nb-InSb NW-Nb quantum device, in which a normal InSb quantum dot (QD) is present between two superconductor Nb-contacted InSb NW segments. Under an applied sufficiently strong magnetic field perpendicular to the substrate and thus to the NW, the two Nb-contacted InSb NW segments turn to become two TS NWs and each hosts a pair of Majorana fermion modes at its ends. Electrical measurements between the two Nb-contacted InSb NW segments in the trivial superconductor phase and in the TS phases are employed to probe the Majorana fermion modes and to study the effect of the interaction between the two Majorana modes located adjacent to the InSb QD. To block the contribution of the supercurrent, we tune the QD to the Coulomb blockade regime. When a sufficiently strong, perpendicular magnetic field is applied to the device, we observe a ZBCP in several consecutive Coulomb blockade diamonds of the QD with both odd and even quasi-particle occupation numbers. Our experiment conclusively rules out the possibility to assign the Kondo physics as a mechanism to the observed ZBCP.

\begin{figure}
\includegraphics[width=8.5cm]{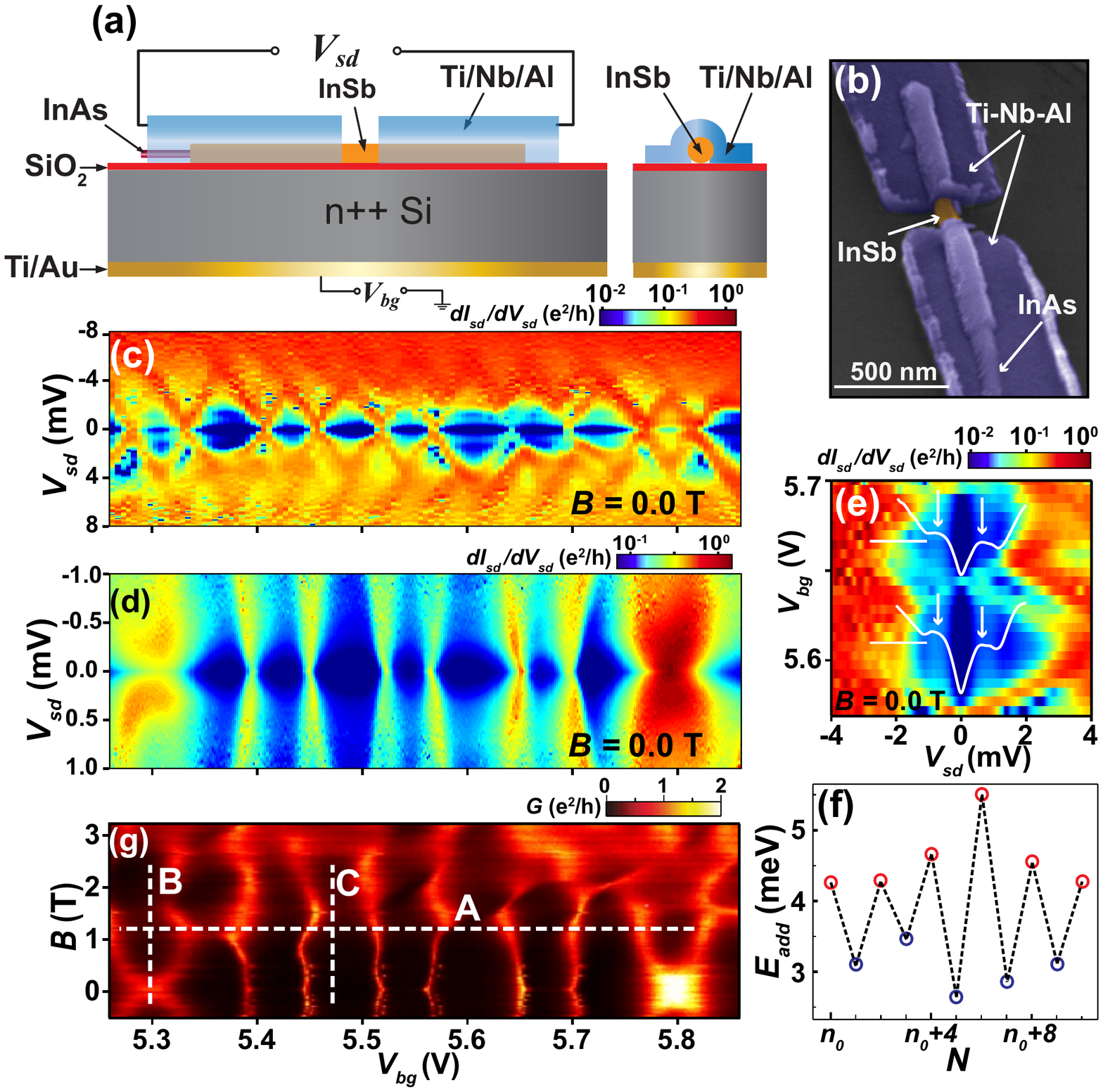}
\caption{\label{fig:1} (a) Schematics showing a side view (left panel) and a cross-section view (right panel) of the structure of a NW based Nb-InSb QD-Nb junction device studied in this work. (b) SEM image of the fabricated Nb-InSb QD-Nb junction device. In this device, the diameter of the InSb NW is about 70 nm, the separation between the two Nb-based contacts is about 150 nm, and the lengths of the InSb NW segments covered by Nb-based contacts are about 780 nm and 720 nm, respectively. (c) Differential conductance on a color scale measured for the device as a function of $V_{sd}$ and $V_{bg}$ (charge stability diagram) at $B=0$~T. (d) Close-up plot of (c) at the low bias voltage region with a higher bias voltage resolution. (e) Zoom-in view of two Coulomb blockade diamonds in the charge stability diagram measurements shown in (c). The two solid curves show the differential conductance plots (line-cuts) at the fixed gate voltages indicated by two short horizontal lines (i.e., the electron-hole symmetry points of the two Coulomb blockade regions). (f) Addition energy, $E_{add}$, of a single quasi-particle to the QD extracted from the measurements in (c) as a function of quasi-particle numbers in the QD. Here, a clear even-odd oscillation behavior of the addition energy can be seen. (g) Linear response conductance on a color scale measured for the device as a function of $V_{bg}$ and the magnetic field $B$. Here, the magnetic field is applied perpendicularly to the substrate and thus to the InSb NW.}
\end{figure}

The structure of our Nb-InSb NW QD-Nb hybrid device is schematically shown in figure \ref{fig:1}a. The device is fabricated from the zincblende InSb part of an epitaxially grown, high crystalline quality, InSb/InAs heterostructure NW~\cite{Caroff2008} on an n-type Si substrate covered by a thin SiO$_2$ layer on top and a Ti/Au back gate electrode on the bottom side (see Supplementary Materials for more details). Two Nb-based superconductor contacts with a separation of 150 nm are defined on the cleaned surface of the InSb NW by electron-beam lithography and an InSb QD is formed at low temperatures between the contacts due to the Schottky barriers formed at the contact-InSb NW interfaces~\cite{NillsonNL2009,NillsonPRL2010}. Figure \ref{fig:1}b shows a scanning electron microscope (SEM) image of the fabricated device. All the electrical measurements reported in this work are performed in a dilution refrigerator at base temperature $T=25$~mK.

The measurements of the fabricated device is first performed for the differential conductance at $B=0$~T as a function of the back gate voltage $V_{bg}$ and the source-drain bias voltage $V_{sd}$ (the charge stability diagram). The results of the measurements are presented on a color scale over a large $V_{sd}$ range in figure \ref{fig:1}c with a zoom-in plot shown in figure~\ref{fig:1}d. Here, clear Coulomb blockade diamond structures can be seen, indicating the formation of the QD between the two superconductor contacts~\cite{NillsonNL2009,NillsonPRL2010}. Through the Coulomb blockade diamond structures, two horizontal high conductance stripes separated by a low conductance gap appear in the small $V_{sd}$ region. This conductance feature is seen more clearly in figure \ref{fig:1}e, where a zoom-in view of a region of two Coulomb blockade diamonds is displayed together with two line-cut plots. 

The high conductance stripes can be attributed to the proximity effect induced superconductivity in the Nb-contacted InSb NW segments with a superconductor energy gap $\Delta^*$. At $V_{sd}=\pm \, 2e\Delta^*$, the quasi-particle tunneling is enhanced by the alignment of the upper (lower) gap edge of the quasi-particle density of states in one Nb-contacted InSb NW segment with the lower (upper) gap edge of the quasi-particle density of states in the other Nb-contacted InSb NW segment. Therefore, $\Delta^* \sim 0.3$ meV at $B=0$ T can be deduced from the measurements~\cite{Deng2012}. The quasi-particle addition energy $E_{add}$ to the QD at $B=0$ T can also be extracted from the measurements and the results are plotted in figure \ref{fig:1}f. It is seen that $E_{add}$ is in the range of 2.7$-$5.6 meV, which is much larger than $\Delta^*$. $E_{add}$ also shows a regular odd-even oscillation behavior and thus the device shows consecutive alternations in the parity of the quasi-particle occupation number in the QD. Figure~\ref{fig:1}g shows the linear response conductance on a color scale measured for the device as a function of $V_{bg}$ and magnetic field $B$ applied perpendicularly to the substrate and thus to the InSb NW at $V_{sd}=4~\mu$V. The conductance peaks (the bright lines) arise from resonance tunneling through the quasi-particle states in the QD. Due to magnetic-field induced energy shifts of these quasi-particle states, these peaks shift in position of $V_{bg}$ as the magnetic field increases. From the low magnetic field region of the measurements, we can also identify the parity of the quasi-particle number in the QD~\cite{NillsonNL2009, NillsonPRL2010}. The results are consistent with the parity extracted from the addition-energy measurements shown in figure \ref{fig:1}f.

In both figures~\ref{fig:1}d and \ref{fig:1}g, two high zero-bias conductance stripes, located in the Coulomb blockade regions at back gate voltages around $V_{bg}=5.3$ V and $V_{bg}=5.8$ V, can be identified at $B=0$~T. These conductance enhancements can be attributed to the spin-1/2 Kondo effect in the Coulomb blockade QD with odd quasi-particle occupation numbers. Note that in a Coulomb blockade region without a clear Kondo effect enhancement, the signatures of Josephson current~\cite{Jogensen2006,Xiang2006}, i.e., sharp and high conductance peaks at zero-bias voltage, are not found in our device. This is because the Josephson effect is strongly suppressed in the system in the presence of the Coulomb blockade effect (see also figure 5 in Supplementary Materials and the corresponding discussion).

\begin{figure}
\includegraphics[width=8.5cm]{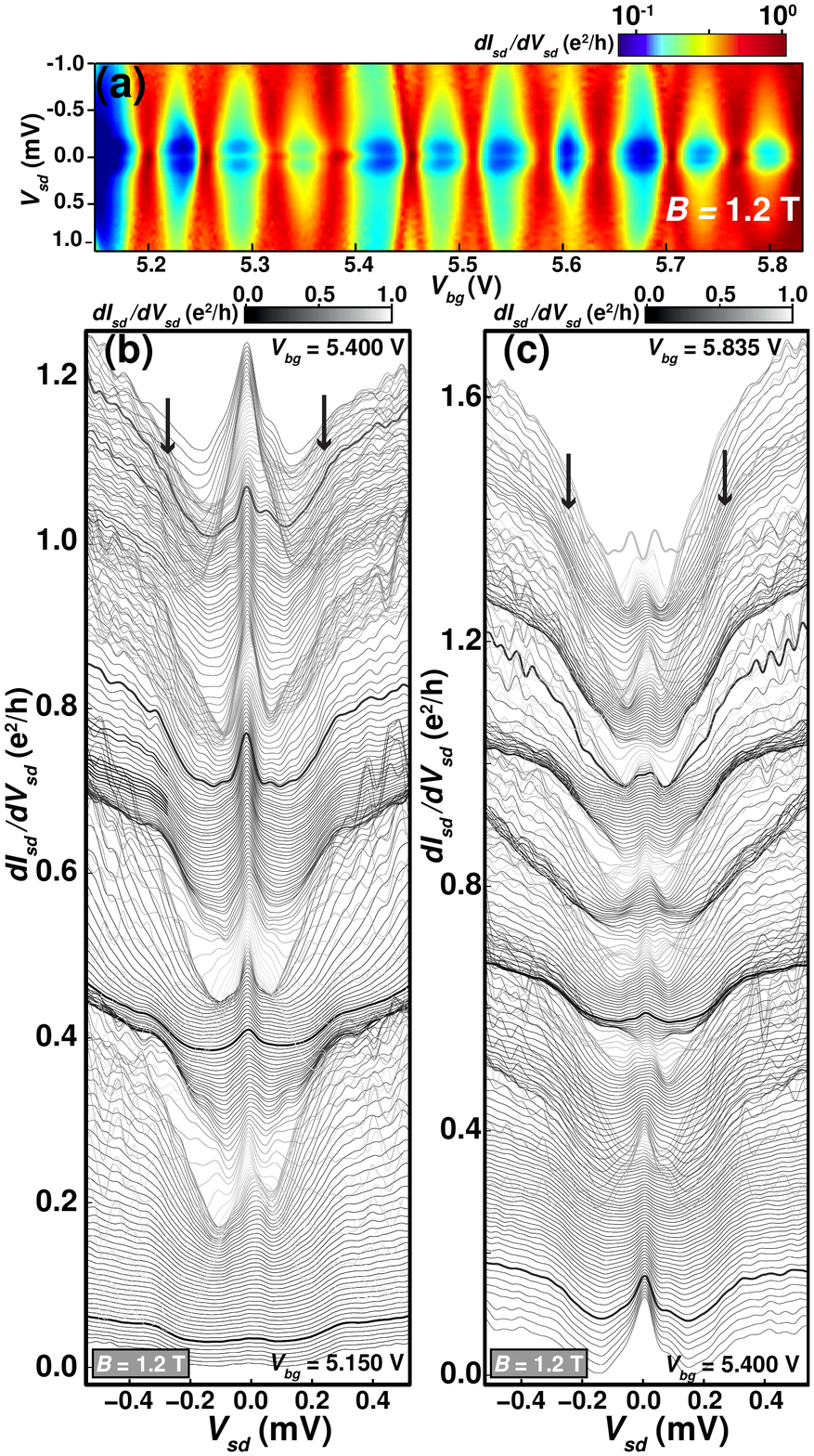}
\caption{\label{fig:2} (a) Charge stability diagram measured for the device at $B=1.2$~T, i.e., along the dashed line A in figure \ref{fig:1}g. It is seen that a stripe of the ZBCP structure appears in every Coulomb blockade diamond found in the figure, irrespective of the quasi-particle occupation number parity in the QD. (b) and (c) Corresponding differential conductance line-cut plots of the measurements at a series of values of $V_{bg}$. The plots shown in panel (b) are the differential conductance measurements at $V_{bg}$ ranging from 5.15 V to 5.4 V, while the plots shown in panel (c) are the differential conductance measurements at $V_{bg}$ ranging from 5.4 V to 5.835 V. For clarity, these line-cut plots are shifted in such a way that values of the differential conductance at $V_{sd} = 0\; \mu$V in these line cuts are successively placed $0.005 e^2/h$ higher than the values of their adjacent, lower gate-voltage line cuts, while their actual zero-bias conductance values are represented by their color in a gray scale. Here, in (ab) and (c), the existence of a ZBCP structure in every Coulomb blockade diamond is more clearly displayed. An overall low differential conductance gap (with its two edges indicated by two arrows on top of each panel) in the low bias voltage region due to the presence of the proximity effect induced superconductor energy gap in the Nb-contacted InSb NW segments and its weak dependence on the back gate voltage $V_{bg}$ can also be identified in the line-cut plots shown in panels (b) and (c).}
\end{figure}

To drive the Nb-contacted InSb NW segments of the device from trivial superconducting phase to nontrivial TS phase, application of an external magnetic field $B$, perpendicular to the Rashba SOI-induced field $B_{\rm{SO}}$, is needed. The magnetic field introduces a Zeeman energy ${E_z} = \frac{1}{2}\left| {g^*} \right|{\mu _B}\tilde{B}$, where ${\mu _B} = {{e\hbar } \mathord{\left/ {\vphantom {{e\hbar } {2{m_e}}}} \right.\kern-\nulldelimiterspace} {2{m_e}}}$ is the Bohr magneton, $g^*$ is the effective $g$-factor, and $\tilde{B}$ is the magnetic field actually applied on the Nb-contacted InSb NW segments, which can be greatly different from $B$ due to the Meissner effect. In general, it is difficult to accurately determine the strength of the externally applied magnetic field $B_T$ at which the phase transition in the Nb-contacted InSb NW segments occurs. For our device, according to the measured properties of the Nb thin film (see Supplementary Materials), the lower limit value of $B_T$ is estimated to be in the range of $0.33\sim0.78$ T.

Figure \ref{fig:2}a shows the charge stability diagram with the differential conductance on a logarithmic color scale measured for the device at $B=1.2$~T, i.e., along dashed line A in figure~\ref{fig:1}g.  Figures \ref{fig:2}b and \ref{fig:2}c are the corresponding line-cut plots of the differential conductance on the linear scale at small source-drain bias voltages. For clarity and to overcome the influence of the background conductance oscillations due to the Coulomb blockade effect, these line-cut plots are offset in such a way that the values of the differential conductance at $V_{sd} = 0\; \mu$V in these line-cuts are successively placed $0.005 e^2/h$ higher than the values of their adjacent, lower gate-voltage line cuts, while their actual zero-bias conductance values are represented by their gray-scale colors. Here, Coulomb diamond structures remain to be clearly seen in the charge stability diagram plot of figure \ref{fig:2}a, indicating the survival of the QD in the device in the presence of the 1.2 T magnetic field. In the line-cut plots of figures \ref{fig:2}b and \ref{fig:2}c, we can see a global conductance gap structure in the small source-drain bias voltage region of $V_{sd}\sim-0.35$ mV to $V_{sd}\sim0.35$ mV, showing that the Nb-contacted InSb NW segments are still in a superconducting state with an energy gap $\Delta^*\sim 0.17$ meV. Strikingly, a pronounced conductance peak appears at zero-bias voltage and goes through the whole $V_{bg}$ region shown in figure~\ref{fig:2}, including both the Coulomb blockade regions and quantum resonance regions. This zero-bias conductance peak (ZBCP) feature is seen in figure \ref{fig:2}a and is even more clearly seen in figures \ref{fig:2}b and \ref{fig:2}c. The ZBCP has a height of up to $\sim0.2~e^2/h$ at the quantum resonances and of up to $\sim0.06~e^2/h$ at the electron-hole symmetry points (i.e., at the centers of the Coulomb blockade diamonds). Moreover, the appearance of the ZBCP is independent of the even-odd parity of the quasi-particle occupation number in the QD, i.e., the ZBCP appears in all the Coulomb blockade diamonds shown in figure \ref{fig:2}, irrespective of whether even or odd the quasi-particle occupation number in the QD is.

\begin{figure}
\includegraphics[width=8.5cm]{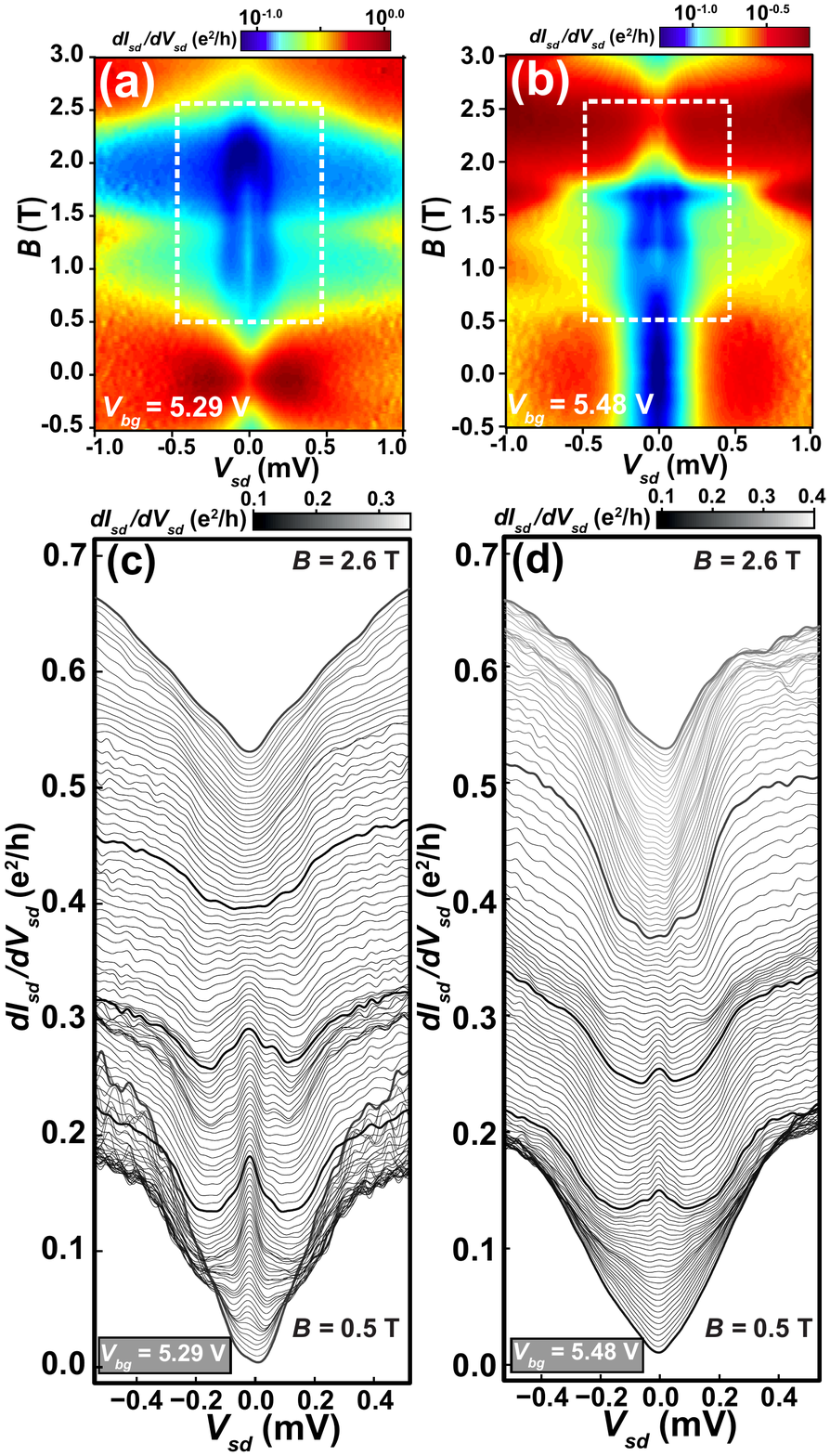}
\caption{\label{fig:3} (a) Charge stability diagrams measured for the device at $V_{bg}=5.29$ V, i.e., along the dashed line B in Fig.~\ref{fig:1}g. (b) Charge stability diagrams measured for the device at $V_{bg}=5.48$ V, i.e., along the dashed line C in Fig.~\ref{fig:1}g. (c) Line-cut plots of the differential conductance measurements in the bias voltage and magnetic field region indicated by the white dashed rectangle in (a). (d) Line-cut plots of differential conductance measurements in the bias voltage and magnetic field region indicated by the white dashed rectangle in (b). For clarity, these line-cuts are shifted in the same way as in figures \ref{fig:2}(b) and \ref{fig:2}(c), i.e., in such a way that values of the differential conductance at $V_{sd} = 0\; \mu$V in these line cuts are placed successively $0.005 e^2/h$ higher than the values of their adjacent, lower gate-voltage line cuts, while their actual zero-bias conductance values are represented by their gray-scale colors. The ZBCP in panels (a) and (c) is seen to emerge at $B\sim 0.6$ T and disappear at $B\sim 1.8$ T, while the ZBCP in panels (b) and (d) is seen to emerge at $B\sim 0.75$ T and disappear $B\sim 2.1$ T. Panels (c) and (d) also show an overall low differential conductance gap in the low bias voltage region due to the presence of the proximity effect induced superconductor energy gap in the Nb-contacted InSb NW segments and its evolution with increasing magnetic field.}
\end{figure}

Figure \ref{fig:3} shows the magnetic field dependent measurements of the differential conductance of the device. Figures \ref{fig:3}a and \ref{fig:3}b display the differential conductance on a logarithmic color scale measured for the device as a function of $V_{sd}$ and magnetic field $B$ perpendicularly applied to the substrate at $V_{bg}=5.29$ V and $V_{bg}=5.48$ V, i.e., along dashed lines B and C in figure \ref{fig:1}g, respectively. The corresponding line-cut plots of the differential conductance are shown in the linear scale in figures \ref{fig:3}c and \ref{fig:3}d. Again, for clarity, these line-cut curves are offset in the same way as in figure~\ref{fig:2}b and \ref{fig:2}c with their actual zero-bias conductance values represented by their gray-scale colors. 

The magnetic field dependent measurements of the differential conductance shown in figures \ref{fig:3}a and \ref{fig:3}c are performed in the Coulomb blockade region of an odd quasi-particle occupation number in the QD, while the measurements shown in figures \ref{fig:3}b and \ref{fig:3}d are performed in the Coulomb blockade region of an even quasi-particle occupation number in the QD (see figures \ref{fig:1}g). Again, from these figures, we can see a $\Delta^*$-induced low-conductance gap and a tendency of decreasing in the width of the gap as $B$ increases. Within the Coulomb blockade regions of both the odd and the even occupation number of quasi-particles in the QD, there is no ZBCP structure when the applied magnetic field is small. However, as the applied magnetic field is increased over a certain value, a ZBCP structure emerges in both Coulomb blockade regions. This ZBCP structure is seen to persist in a finite range of magnetic fields before it finally disappears at higher magnetic fields--it emerges at $B\sim0.6$ T and disappears at $B\sim1.8$ T in figures \ref{fig:3}a and \ref{fig:3}c, while in figures \ref{fig:3}b and \ref{fig:3}d it emerges at $B\sim 0.75$ T and disappears at $B\sim 2.1$ T. 

Similar results as in figures \ref{fig:2} and \ref{fig:3} have also been observed in the charge transport measurements of a different NW based Nb-InSb QD-Nb hybrid device (see Supplementary Materials). With the presence of several Coulomb blockade diamonds in the charge stability diagram measurements of the devices, we have showed for the first time that the ZBCP structure is independent of the even-odd parity of quasi-particle occupation numbers in the QD. The parity independence favors the assignment of the ZBCP structure to the MF physics. At a sufficiently strong applied magnetic field and a suitable gate voltage, it is possible to drive the two Nb-covered InSb NW segments in such a device to TS phase, leaving the intermediate QD to remain as a trivial object. Hence, two pairs of MF bound states, spatially separated by the QD, can be created in the TS-QD-TS hybrid system. However, in the presence of a finite coupling between the two TS NW segments, the two MF bound states located adjacent to the QD (i.e., the inner two MF bound states) can interact and hybridize into a pair of quasi-particle states with finite energies. The other pair of MF bound states (i.e., the outer two MF bound states) located at the two ends of the entire InSb NW remain at zero energy and thus the entire system, including the two Nb-contacted NW segments and the QD, behaves as a nontrivial TS NW (see the Supplementary Materials in Ref.~\cite{Deng2012}). In our experiment, it is this outer pair of MF states that make Cooper pair transport between the two Nb contacts possible, leading to an enhancement of the conductance at zero-bias voltage. Because the existence of MF bound states in the TS-QD-TS system is independent of the parity of the quasi-particle occupation number as well as the energy position of the quasi-particle states in the QD, the ZBCP can go through more than ten consecutive Coulomb blockade diamonds, regardless of the parity of quasi-particle occupation numbers and the energy position of the quasi-particle states in the QD.

\begin{figure}[th]
\includegraphics[width=8.5cm]{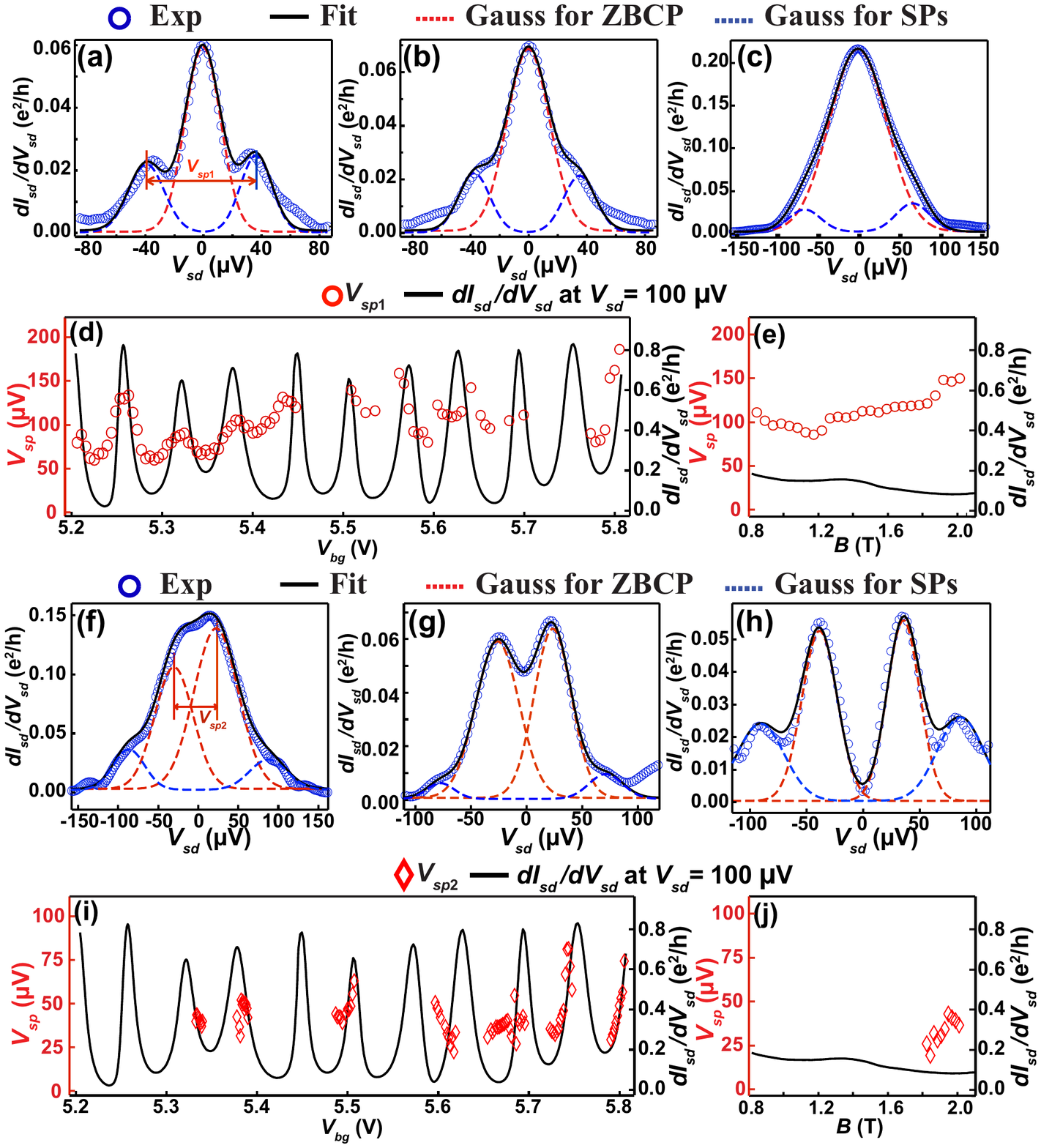}
\caption{\label{fig:4} (a)-(c) Differential conductance as a function of $V_{sd}$ measured at $B = 1.2$ T for the device at $V_{bg}=5.385$ V, 5.395 V, and 5.445 V, respectively. Blue circles are the experimental data, red dashed lines are the Gaussian fits to the ZBCPs, blue dashed lines are the Gaussian fits to the side peaks, and black solid lines are the overall fitting results to the experimental data obtained by the sum of the Gaussian fits to the ZBCP and to the side peaks. (d) Separation (red circles) of the two side peaks $V_{sp1}$, as defined in panel (a), extracted from figures \ref{fig:2}b and \ref{fig:2}c as a function of $V_{bg}$. The solid line in the panel shows the differential conductance $dI_{sd}/dV_{sd}$ measured for the device at $V_{sd} = 100\;~\mu$V. Here, it is seen that the $V_{sp1}$ oscillations and the differential conductance oscillations are closely correlated. (e) Value of $V_{sp1}$ (red circles) extracted from figure \ref{fig:3}(c) as a function of $B$ plotted together with the differential conductance (solid line) measured for the device at $V_{sd} = 100\;~\mu$V. (f)-(h) Differential conductance as a function of $V_{sd}$ measured at $B = 1.2$ T for the device at $V_{bg}=5.765$ V, 5.815 V and 5.835 V, respectively. Here the ZBCP is seen to split into two peaks. Again, blue circles are the experimental data and blue dashed lines are the Gaussian fits to the side peaks. But red dashed lines are the Gaussian fits to the split peaks from the ZBCP. Black solid lines are the overall fitting results to the experimental data obtained by the sum of the Gaussian fits to the peaks split from the ZBCP and to the side peaks. (i)  Splitting of the ZBCP $V_{sp2}$ (red circles), as defined in panel (f), extracted from figures \ref{fig:2}b and \ref{fig:2}c as a function of $V_{bg}$ and the differential conductance (solid line) measured for the device at $V_{sd} = 100\;~\mu$V. (j) Value of $V_{sp2}$ (red circles) extracted from figure \ref{fig:3}(c) as a function of $B$ together with the differential conductance (solid line) measured for the device at $V_{sd} = 100\;~\mu$V.}
\end{figure}

For a Josephson QD junction, there are several other possible mechanisms that can lead to a ZBCP, such as the Kondo effect~\cite{Lee2012}, quantum phase transition (QPT)~\cite{Lee2013}, and Josephson supercurrent~\cite{Jogensen2006, Doh05,Xiang2006,Dam06,Deon2011}. The Kondo effect depends strongly on whether the ground state of the QD is spin-singlet or spin-doublet and thereby depends on the parity of the quasi-particle occupation number in the QD, in clear contrast to the observed parity independence of the ZBCP in our experiment. In addition, the Kondo effect (including exotic Kondo effects~\cite{SchmidPRL2000, SasakiNature2000, JarilloNature2005}) induced ZBCP will split into two conductance peaks located at finite bias voltages in finite magnetic fields, which is especially true for the InSb NW QD with a giant $g^*$-factor~\cite{NillsonPRL2010}. However, the observed ZBCP in our system stays at the zero bias voltage for a magnetic field range of $>1$ T (corresponding to a Zeeman energy of $V_Z\sim 2$ to $4$ meV in the InSb NW) (figures \ref{fig:3}). Hence, it is not consistent to attribute the ZBCP observed in our experiment to the Kondo effect. A magnetic field induced QPT from a singlet to a spin-polarized ground state or from a spin-polarized to a singlet ground state in the QD could also lead to a ZBCP~\cite{Lee2013}. However, such a QPT is sensitively dependent on the energy position of the quasi-particle state in the QD and spin-resolved Andreev levels and, thus, can only occurs at certain values of $V_{bg}$ and $B$. Our experiment shows that the observed ZBCP appears continuously over a large range of $V_{bg}$ (covering several Coulomb blockade diamond regions) and a large range of $B$ ($>1$ T, see the discussion above), which does not resemble the characteristic behavior of the ZBCP derived from the QPT.  In order to rule out the possibilities to associate the Josephson supercurrent to the observed ZBCP, we have in the experiment tuned the QD into well-defined Coulomb blockade regions. For a trivial superconductor system, the Cooper pair transport through the QD in such a Coulomb blockade region can be suppressed and no supercurrent induced conductance peak could be observed (see also figure 5 in Supplementary Materials and the corresponding discussion). However, in strong contrast, our measurements presented in figure~\ref{fig:2} show that there exist a ZBCP in several entire Coulomb blockade diamond regions. Thus, the ZBCP feature existing in the entire gate voltage region shown in figure~\ref{fig:2} could not be attributed to the supercurrent mechanism.

As we discussed previously, two coherently connected MFs will hybridize into a pair of quasi-particle states with finite energies (see also the Supporting Information in Ref.~\cite{Deng2012}). This hybridization will lead to splitting of the ZBCP and can serve as an important signature of the Majorana physics. In our TS-QD-TS system, there would exist two pairs of zero-energy MFs when there were no coupling between the two TS NW segments. In reality, the two inner MFs can be coherently coupled through the QD, leading to the creation of a pair of quasi-particle states at finite energies, while the outer two MFs can remain intact and staying at zero energy. As a consequence, the transport measurements can show a triple conductance peak structure, with two side differential conductance peaks appearing at finite bias voltages tunable by tuning the quasi-particle states in the QD and with the middle peak still staying at the zero bias voltage irrespective of the energy positions of the quasi-particle states in the QD. 

This triple conductance peak structure has indeed been observed in the measurements shown in figures~\ref{fig:2} and \ref{fig:3}. Figures~\ref{fig:4}a-\ref{fig:4}c show the results of the measurements (open circles) at three selected back-gate voltages of $V_{bg}=5.385$, 5.395, and 5.765 V, displaying the characteristics of the triple conductance peak structure. In each of these figures, the red dashed line and the two blue dashed lines are the Gaussian fits to  the ZBCP and the two side conductance peaks, while the solid black line represents the actual result of the fitting, i.e., the sum of the three individual Gaussian fits. It can be seen in figures~\ref{fig:4}a-\ref{fig:4}c that, in all the three cases, the experimental data are well reproduced by the fits.

It is clear that the bias voltage positions of the two side conductance peaks are $V_{bg}$-dependent. To investigate the behavior of the two side conductance peaks quantitatively, we define $V_{sp1}$ as the distance in $V_{sd}$ between the two side peaks (see the definition in figure \ref{fig:4}a) and  plot $V_{sp1}$ extracted from figures \ref{fig:2}b and \ref{fig:2}c as a function of $V_{bg}$ in figure \ref{fig:4}d (red circles). It is seen that $V_{sp1}$ shows regular oscillations with amplitudes in a range of $50-150$ $\mu$V. The black solid line in figure~\ref{fig:4}d shows the differential conductance measured at $V_{sd}=100$ $\mu$V as a function of $V_{bg}$, displaying the regular Coulomb conductance oscillations. Evidently, the $V_{sp1}$ oscillations are closely correlated to the Coulomb conductance oscillations in the transport through the QD. 

This oscillation correlation phenomenon is consistent with our hypothesis that the hybridization of the two inner MFs are strongly influenced by the quasi-particle states in the QD. When the energy of a quasi-particle state in the QD is aligned with the MF states in the TS NWs, the interaction between the two inner MFs is enhanced via the quasi-particle state, leading to a large separation between the two quasi-particle states and thus a large value of $V_{sp1}$. However, as the quasi-particle state in the QD is moved away in energy from the MF states in the TS NWs as at the electron-hole symmetry point of a Coulomb blockade diamond region, the interaction between the two inner MFs is reduced and thus a small value of $V_{sp1}$ is observed.

Figure \ref{fig:4}e shows $V_{sp1}$ extracted from the measurements presented in figure \ref{fig:3}c as a function of the magnetic field and the corresponding differential conductance measured at $V_{sd}=100$ $\mu$V. Here, we see that $V_{sp1}$ shows smooth variation with increasing  magnetic field, in contrast to the clear oscillations of $V_{sp1}$ with increasing $V_{bg}$ seen in figure \ref{fig:4}d. This is because the device remains in a deep Coulomb blockade region in the whole magnetic field range in figure \ref{fig:4}e and there is no on-off resonance alternation induced modulation on the interaction between the two inner MFs.

Finally, it is worthwhile to note that small splittings of the ZBCP have also been observed in the measurements of our device shown in figures~\ref{fig:2} and \ref{fig:3} at certain back gate voltages and magnetic fields. Figures~\ref{fig:4}f-\ref{fig:4}h show the results of the measurements (open circles) at $V_{bg}=5.765$, 5.815, and 5.835 V. Here, it is clearly seen that the ZBCP splits into two peaks which, together with the two side differential conductance peaks, form a quadruple conductance peak structure.  The quadruple conductance peak structure can also be well fitted by Gaussians as shown by the dashed lines in figures \ref{fig:4}f-\ref{fig:4}h. Similarly as for $V_{sp1}$, we can define $V_{sp2}$ as the distance in $V_{sd}$ between the peaks split from the ZBCP. Figure \ref{fig:4}i shows the value of $V_{sp2}$ (red circles) extracted from the measurements presented in figures \ref{fig:2}b and \ref{fig:2}c as a function of $V_{bg}$. However, in contrast to $V_{sp1}$, $V_{sp2}$ does not show clear correlated oscillations with Coulomb conductance oscillations (the black solid line in figure \ref{fig:4}i). It is also observed that the splitting of the ZBCP does not occur at all the resonances and it does not only occur at the resonances either. In particular, the splitting of the ZBCP is seen to occur even in deep Coulomb blockade regions. Figure \ref{fig:4}j shows $V_{sp2}$ extracted from the measurements presented in figure \ref{fig:3}c as a function of magnetic field and the corresponding differential conductance measured at $V_{sd}=100$ $\mu$V. Here, the splitting of the ZBCP is clearly seen in the high magnetic field region. However, the ZBCP does not show any visible splitting at any magnetic fields in the magnetic field dependent measurements shown in figure \ref{fig:3}d. 

Splitting of the ZBCP has recently been predicted as an evidence for the existence of the Majorana modes in a TS NW~\cite{SarmaPRB2012}. Based on this prediction, we can attribute the splitting of the ZBCP observed in our experiment to hybridization of the two outer MFs. Similarly as for the two inner MFs in our system, the interaction between the outer two MFs is, in principle, influenced by the chemical potential, Zeeman energy and the quasi-particle states in the QD. However, in comparison with the two inner MFs, the interaction between the two outer MFs is extremely weak and the splitting of the ZBCP is thus much smaller as it is seen in figure \ref{fig:4}i-\ref{fig:4}j.

In summary, we have studied a Nb-InSb NW QD-Nb hybrid device made from an epitaxially grown InSb NW with strong SOI on a Si/SiO$_2$ substrate by charge transport measurements. At zero magnetic field, the device shows a series of well defined Coulomb blockade diamonds and the Kondo effect. At a fixed but sufficiently strong magnetic field applied perpendicularly to the substrate and thus to the NW, a pronounced ZBCP structure is observed in the Coulomb blockade regions and is found to be present in more than ten consecutive Coulomb blockade diamonds, irrespective of the even-odd parity of the quasi-particle occupation number and of the energy position of the quasi-particle states in the QD. We have also observed that the ZBCP is in most cases accompanied by two side differential conductance peaks located at finite bias voltages, forming a triple conductance peak structure. The splitting of the two side peaks is found to be correlated to the background conductance of the device. These observations are consistent with the signatures of MF physics in the device: In a NW based TS-QD-TS system, the two inner MFs are coherently coupled via the QD and are hybridized into a pair of quasi-particles with finite energies, while the two outer MFs remain as zero-energy modes and the entire system behaves as a TS NW.

We acknowledge the financial supports from the Swedish Research Council (VR), the National Basic Research Program of the Ministry of Science and Technology of China (Nos. 2012CB932703 and 2012CB932700) and the National Natural Science Foundation of China (Nos. 91221202 and KDB201400005).

\end{document}